# Variational Principles of Physics and the Infinite Ramsey Theory

Edward Bormashenko*

[a]Department of Chemical Engineering, Ariel University, Ariel, POB 3, 407000, Israel

*Correspondence to: Edward Bormashenko edward@ariel.ac.il

**Abstract**

Application of the Ramsey Infinite Theorem to the variational principles of physics is discussed. The Hamilton Least Action Principle states that, for a true/actual trajectory of a system, Hamilton's Action is stationary for the trajectories, which evolve from the fixed initial space-time point to the fixed final space-time point. The Hamilton Principle distinguishes between the actual and trial/test trajectories of the system in the configurational space. This enables the transformation of the infinite set of points of the configurational space (available for the system) into the bi-colored, infinite, complete, graph, when the points of the configurational space are seen as the vertices, actual paths connecting the vertices are colored with red; whereas, the trial paths are colored with green. According to the Ramsey Infinite Theorem, there exists the infinite, monochromatic chain of the pathways/clique, which are completely built from actual or virtual paths, connecting the intermediate states of the system. The same procedure is applicable to the Maupertuis's principle (classical and quantum), Hilbert-Einstein relativistic variational principle and reciprocal variational principles.

## 1. Introduction

Variational Principles have a long and distinguished history in Mathematics and Physics [1-3]. Apart from general formulations of physical principles, equivalent to local differential equations, they also are useful to approximate problems too difficult for analytic solutions [1]. In the past century quantum variational calculations have been ubiquitous as an approximate method for the ground state of many complicated systems [1]. The laws of physics in there most general form are expressed mathematically as variational principles [4-5]. These principles possess two main characteristics. First, they appear to be universal. Second, they express physical laws as the results of optimal equilibrium conditions between conflicting causes [2]. In other words, they present natural phenomena as problems of optimization under preset constraints [2]. This fundamental idea in modern physics is rooted in the Fermat studies and his least time principle in optics. Rigorously speaking the Fermat Principle was anticipated nearly a millennium ago by the Arab scientist Ibn al-Haytham and it was inspired by the reflections of the Greek polymath Hero of Alexandria (Ηρων ο Αλεξανδρε'υς) on light almost two millennia ago [6-8]. These ideas were further developed in the framework of the calculus of variations of Euler and Lagrange [4-5]. In 1844, Maupertuis revealed, with the help of Euler, the least action principle in mechanics [2]. The comprehensive review of the state of art in the field of the modern understanding of variational principles of physics is found in ref. 1. They serve as basic axiomatic foundations of mechanics (classical and quantum [9-11]), field theory [10] and even thermodynamics [12-13].

The present paper reports the re-shaping of the variational principles of physics, carried out within the framework of the infinite Ramsey theory [14-16]. The Ramsey theory, introduced by the British mathematician and philosopher F. P. Ramsey, is the field of combinatorics/graph theory that deals with a specific kind of mathematical structure, namely complete graphs [14-16]. A graph is a mathematical structure comprising a set of objects in which some pairs of objects are in some sense "related" [14-16]. Finite and infinite Ramsey theories were developed. Let us start from the finite Ramsey theory. The Ramsey theorem states that sufficiently large, finitely colored, complete graphs must contain a specific monochromatic sub-graph. The typical problem considered by the Ramsey theory is the so-called "conference problem", which predicts the minimum number $R(m,n)$ of the conference participants (each of whom is either a friend or a stranger to the others) must be invited so that at least $m$ of the participants will be acquainted with each other, or at least $n$ of them will not be familiar with each other [14-16]. In this case $R(m,n)$ is defined as a Ramsey number [14-16]. Consider the particular formulation of the conference problem: "What is the smallest number of people in a gathering, every two of whom are either friends or strangers, that will guarantee that there are either three mutual friends or three mutual strangers in the conference room"? In this particular case $R(3,3) = 6$ [14-16].

The infinite Ramsey theorem, in turn, is formulated as follows: Let $K_\omega$ denote the complete graph on the vertex set $N$. For every $c > 0$, if we color the edges of $K_\omega$ with $c$ colors, then there must be an infinite monochromatic clique. A clique of the undirected graph is a subset of vertices of the graph every two distinct vertices in the clique are adjacent. We demonstrate, how synthesis of the Ramsey Theory and variational principles of physics is possible.

2. **The Hamilton Principle as it is Seen From the Point of View of the Infinite Ramsey Theory**

The Hamilton Least Action Principle addresses the Hamilton action denoted $S$, which is defined as an integral along any actual or virtual (conceivable or trial) space-time trajectory $q(t)$ connecting two specified space-time events/points, which are the initial event/point $A \equiv (q_A, t_A)$ and the eventual event/point $B \equiv (q_B, t_B)$, where $q(t)$ is the generalized coordinate.

$$S = \int_{t_A}^{t_B} L(q(t), \dot{q}(t)) dt, \qquad (1)$$

where $\dot{q}(t) \equiv \frac{dq(t)}{dt}$ is the generalized velocity, and $t$ is time. Generally speaking, $q$ denotes the complete set of independent generalized coordinates $q_1, q_2 \ldots q_p$, , where $p$ is the number of degrees of freedom of the physical system [1-3]. For a sake of simplicity, only holonomic generalized coordinates are taken into account [1-3]. It should be emphasized, that the initial and final points are fixed in the space of generalized coordinates, which is called the configurational space [3].

The Hamilton Least Action Principle states that, for a true/actual trajectory of a system, Hamilton's Action $S$ is stationary for the trajectories, which evolve from the fixed initial space-time point $A \equiv (q_A, t_A)$ to the fixed final space-time point $B \equiv (q_B, t_B)$ [1-3]. In other words Eq. 2 takes place [1-3]:

$$(\delta S)_{t_B - t_A} = 0 \qquad (2)$$

The Hamilton Principle distinguishes between the actual and trial/test trajectories of the system in the configurational space. Let us illustrate this distinction with the sketch depicted in **Figure 1**. Consider two points $A \equiv (q_A, t_A)$ and $B \equiv (q_B, t_B)$ in the configurational space of the given physical system. The evolution of the system from $A \equiv (q_A, t_A)$ to $B \equiv (q_B, t_B)$ may occur according two pathways, namely: "actual" depicted in **Figure 1** with a red link (in this case points $A \equiv (q_A, t_A)$ and $B \equiv (q_B, t_B)$ are "friends", and trial/virtual (which does not actually take place), shown in **Figure 1** with a green link (in this case points $A \equiv (q_A, t_A)$ and $B \equiv (q_B, t_B)$ are

"strangers" in the terms of the Ramsey Theory [14-16]). We keep the proposed bi-colored procedure throughout the text. It should be emphasized, that the sketch, depicted in **Figure 1** is not a graph; in the "true graph" two points/vertices are connected with a single link/edge [14-16].

Now consider the infinite number of points $q_1, q_2 \ldots q_N \ldots$ in the configurational space of the physical system (again every point denotes the complete set of independent generalized coordinates). The motion of the system is desribed by the functions $q_1(t), q_2(t) \ldots q_p(t), p$ – is the finite numbers degrees of freedom of the system. We consider only the points which accomplish with the actual physical constraints implied on the system: for example, for the mathmatical pendulum, these points are restricted by the fixed length of the pendulum. Every point (seen as vertice of the graph) is joint by a link corresponding to actual/red or vitual/green link with the another fixed point of the avaiable configurational space, as shown in **Figure 2**. According to the introduced coloring procedure the red link corresponds to the actual trajectory of the system in the configurational space, $\delta S = 0$ is true for the red link (in other words functions $q_1(t), q_2(t) \ldots q_p(t)$ accomplish the Lagrange equations of the motion of the given system). A green link, in turn, corresponds to the trial/virtual path, which does not actually occurs, $\delta S \neq 0$ takes place for the red links. Two points of the configurational space are connected with a single link. Thus, the complete graph, emerges. According to the Infinite Ramsey Theorem, infinite, complete monochromatic subgraph (either green or red) will necessarily appear in the graph, in other words an infinite monochromatic clique will be present in the graph [14-16]. Thus, the actual or trial/virtual trajectory connecting the points in the configurational space will be necessarily present in the graph. The equations of classical mechanics are time-reversible; thus the. addressed infinite bi-colored complete graph will be non-directional. It should be emphasized that that the suggested procedure provides the transition from the continuous to discrete description of the system.

Thus, we demonstrated the following theorem: for any infinitely countable set of points of the configurational space $q_1, q_2 \ldots q_N \ldots$ available for a physical system there exists the infinite, monochromatic chain of the pathways/clique, which are completely built from actual or virtual paths, connecting the intermediate states of the system. It should be emphasized that the Infinite Ramsey Theory says nothing about the kind of the infinite monochormatic graph: it remains unclear, actual or virtual pathway will appear.

### 3. Maupertuis' principle and Infinite Ramsey Theory

Consider the configurational space of the physical system $q_i$, $i = 1 \ldots p$ , $p$ – is the finite numbers degrees of freedom of the system. Now we assume that the Lagrangian $L(q, \dot{q})$ and therefore the Hamiltonian $H(p, q)$ do not involve the time explicitly; so that energy of the system is conserved [9]. In this case, Maupertuis' variational principle is applicable [9]. Let us introduce the value which is called the "abbreviated action" and denoted $S_0$:

$$S_0 = \int \sum_i p_i \delta q_i \qquad (3)$$

where $p_i = \frac{\partial}{\partial \dot{q}_i} L\left(q, \frac{dq}{dt}\right)$ are the generalized momenta of the system. Maupertuis' variational principle states that the actual motion of the system satisfies Eq. 4:

$$\delta S_0 = 0 \qquad (4)$$

In other words, the abbreviated action has a minimum with respect to all paths, which satisfy the law of conversation of energy and path through the final point at any instant [9]. Jacobi re-formulated the Maupertuis principle as follows:

$$\delta \int \sqrt{2m(E - U)} dl = 0 , \qquad (5)$$

where $U$ is the potential energy of the system, the integral is taken between two given points in space [9]. When compared to the Hamilton principle Maupertuis's principle: i) uses the abbreviated action integral over the generalized coordinates, varied along all constant energy paths ending at $A \equiv (q_A)$ and $B \equiv (q_B)$, ii) determines only the shape of the trajectory in the generalized coordinates, and does not supply the exact functions $q_i(t)$, iii) requires that the two endpoint states $A \equiv (q_A)$ and $B \equiv (q_B)$ be given and that energy be conserved along every trajectory. With all these restrictions considered, we apply now the Ramsey bi-coloring procedure introduced in Section 2. Consider the infinite number of points $q_1, q_2 \ldots q_N \ldots$ in the available configurational space of the physical system. We connect two points with the red link, when the condition $\delta S_0 = 0$ is fulfilled for that path connecting the points, or connect the points with the green link when $\delta S_0 \neq 0$ is true for the points. Thus, the red link corresponds to the actual shape of the trajectory of the motion, and the points connected with the red links are "friends" within the Ramsey Theory terminology; whereas, the green link, in turn, corresponds to the trial shape of the trajectory, which actually does occur, and these points, are correspondingly "strangers". Two points of the configurational space are connected with a single link. Thus, a complete, infinite, bi-colored, Ramsey, source graph emerges. Again, according to the Infinite Ramsey Theorem, infinite, complete monochromatic subgraph will appear in the source graph, in other words an infinite monochromatic clique will be necessarily present in the graph. Thus, we demonstrated the following theorem, representing the Ramsey interpretation of the Maupertuis's principle: consider the system in which the energy is conserved. For any infinitely countable set of points of the configurational space $q_1, q_2 \ldots q_N$ available for the system there exists the infinite sequence of trajectories, which are completely built from actual or virtual/trial trajectories, connecting the intermediate states of the system. Again, Ramsey Theory says nothing about what kind of the infinite monochromatic graph/clique: actual or virtual will be present.

The aforementioned Ramsey interpretation of the Maupertuis's principle enables straightforward quantum mechanics extension. The time-dependent Schrödinger equation may be derived from the quantum variational principle, which is formulated as follows [17-19]:

$$\delta S_q = \delta \int \psi^* (\widehat{H} - E)\psi dq = 0, \qquad (6)$$

where $S_q$ is a quantum action, $\psi$ is a wave function, $\widehat{H}$ is the Hamiltonian operator, and $E$ is the energy of the system. Eq. 6 is equivalent to the Schrödinger variational principle [17-18]:

$$\delta \left( \frac{\langle \psi \widehat{H} \psi \rangle}{\langle \langle \psi | \psi \rangle \rangle} \right) = 0 \qquad (7)$$

It should be emphasized, that within the quantum variational principles, a wave function $\psi$ is varied. Thus, the following Ramsey re-interpretation of the quantum variational principles, given by Eqs. 6-7, emerges: consider the infinite number of points $q_1, q_2 \ldots q_N \ldots$ in the available configurational space of the quantum system. We connect two points with the red link, when the condition $\delta S_q = 0$ is fulfilled for that wave function connecting the points, or, otherwise, connect the points with the green link when $\delta S_q \neq 0$ is true for the trial wave function. Thus, the red link corresponds to the actual wave function, and the points connected with the red links are "friends"; whereas, the green link, in turn, corresponds to the trial wave function, and these points, are correspondingly "strangers". Two points of the configurational space are connected with a single link. Thus, a complete, infinite, bi-colored, Ramsey, source graph emerges. This graph will necessarily contain an infinite monochromatic clique, built of the sequence of actual or trial wave functions $\psi$.

## 4. Hilbert-Einstein Relativistic Variational Principle and the Infinite Ramsey Theory

The Hilbert-Einstein relativistic action $S_r$ is the action that yields the Einstein field equations through the stationary-action principle. The classical and original Hilbert-Einstein action is defined as follows:

$$S_r = \int_{Q^4} L_{HE}(g) d\Omega, \qquad (8)$$

where $g_{\mu\nu}$ is the space-time metric 4-tensor, $d\Omega = d^4 r \delta \sqrt{-|g(r)|}$ is the invariant 4-volume element of the Riemann space-time $\{Q^4, g(r)\}$, with $d^4 r = \prod_{i=0...3} dr^i$ is canonical measure of $Q^4$, $|g(r)|$ is the determinant of $g(r)$ and $L_{HE}$ is the Einstein-Hilbert Lagrangian [3, 10]. According to the classical Hilbert-Einstein approach action depends only on the gravitational field $g(r) = g_{\mu\nu}$, whose independent 4-tensor components represent the generalized Lagrangian coordinates. The movement of a free particle $m$ in the general relativity fulfils the demands of the principle of the least action:

$$\delta S_r = -mc\delta \int ds = 0, \qquad (9)$$

where $-ds^2 = g_{\mu\nu} dx_\mu dx_\nu$. A free particle moves in such a way that $\delta S_r = 0$ and its word line is an extremal connecting two fixed points in four-dimensional space-time. Now we apply the procedure introduced in Section 2, namely we distinguish between actual and virtual/trial paths of the particle/system in the Riemann space-time $\{Q^4, g(r)\}$. We consider the infinite number of points in the space-time. "Actual" word lines connecting the points we color with a red link, and the trial/virtual world lines we connect with a green link. Now the application of the Infinite Ramsey Theory is straightforward. For a sake of simplicity, consider the motion of a free relativistic particle. For any infinitely countable set of points of the Riemann space-time there exists the infinite mono-colored chain of the world lines, which are completely built from actual or virtual world lines paths, connecting the intermediate locations of the particle in the four-dimensional space-time. Again, the Ramsey Infinite Theory supplies the convenient framework for the "quantization" of the continuous motion of the relativistic particle.

Limitations of the classical Hilbert-Einstein variational relativistic principle are discussed in ref. 20. The authors demonstrated that only so-called unconstrained variational framework correctly reproduces the Einstein Field Equations as extremal equations [20]. However, the extension of the Infinite Ramsey Approach to this variational framework is straightforward.

### 5. Reciprocal Variational Principles and the Infinite Ramsey Theorem

The authors of refs. 17-18 demonstrated the so-called reciprocal variational principles of mechanics. For example they proved that the Maupertuis' variational principle, given by Eqs. 3-4 may be re-shaped as follows:

$$(\delta \bar{E})_{S_0} = 0 , \qquad (10)$$

where $\bar{E}$ is the mean energy of the system and $S_0$ is the abbreviated action given by Eq. 3 or alternatively by Eq. 5 [17-18]. Application of the Ramsey Infinite Theorem to the reciprocal Maupertuis' variational principle follows the procedure discussed in detail in Section 3.

**Conclusions**

Variational principles of physics are considered as the most general laws forming the basis of mechanics (classical, quantum and relativistic), field theory, optics and thermodynamics [9-13, 21]. These principles are universal and they express physical laws as the results of optimal equilibrium conditions between conflicting causes. We demonstrate how the synthesis of the variational principles of physics and the Infinite Ramsey Theory is possible. Variational principles of physics distinguish between the actual and trial/virtual trajectories, being passed by a physical system in the pre-established space, which may be the configurational space or the Riemann space-time in the relativistic mechanics. The points of the pre-established space are seen now, as the vertices of the graph, and the actual/trial pathways are seen as the links connecting the vertices, which form the countably infinite set. Thus, the bi-coloring of the emerging infinite graph becomes possible, when actual pathways accomplishing the appropriate variational principles are colored with red, and the trial pathways are colored with green. When the infinite, countable set of points/vertices is considered, the complete, infinite, bi-colored Ramsey graph emerges. The procedure may be implemented for the Hamilton, Maupertuis, Hilbert-Einstein or Fermat variational principles [17, 18, 21]. According to the Ramsey Infinite Theorem, there exists the infinite, monochromatic chain of the pathways/clique, which is completely built of the actual or virtual paths, connecting the intermediate states of the system. Consider, for example, the quantum Maupertuis principle. In this case, the red link corresponds to the actual wave function, and the points connected with the red links are "friends"; whereas, the green link, corresponds to the trial wave function, and these points, are correspondingly "strangers". Two points of the configurational space are connected with a single link. Thus, a complete, infinite, bi-colored, Ramsey graph emerges. This graph will necessarily contain an infinite monochromatic clique, built of the infinite sequence of actual or trial wave functions $\psi$. Regrettably, the Ramsey Theory gives no indication: what kind of the sequence/clique (actual or trial) will be present in the graph. The extension of the suggested procedure to the relativistic or reciprocal variational principles is straightforward [17-18]. Involving the Ramsey Infinite Theory enables the "discretization" of the variational principles of physics.

AKNOWLEDGEMENTS

The author is indebted to Dr. Nir Shvalb for useful discussions.

**References**

[1] C. G. Gray, G. Karl, V. A, Novikov, Progress in classical and quantum variational principles, Rep. Prog. Phys. 2004, 67, 159.

[2] J.-L. Basdevant, Variational Principles in Physics, 2nd Ed., Springer-Nature, Cham, Switzerland, 2023.

[3] C. Lanczos, The Variational Principles of Mechanics, Dover Publications, NY, USA, 1970.

[4] I. M. Gelfand, S. V. Fomin, Calculus of variations. New York, USA, Dover, 2000.

[5] M. A. Kot, First course in the calculus of variations. New York, NY, USA, American Mathematical Society; 2014.

[6] M. Ceccarelli, Distinguished Figures in Mechanism and Machine Science: Their Contributions and Legacies; Springer: Dodrecht, The Netherlands, 2007; pp. 217–247.

[7] P. Adamson, Philosophy in the Islamic World: A History of Philosophy without Any Gaps; Oxford University Press: Oxford, UK, 2016; p. 77.

[8] M. Frenkel, S. Shoval, Ed. Bormashenko, Materials 2023, 16(24), 7571.

[9] L. D. Landau, E. M. Lifshitz, Mechanics. Vol. 1 (3rd ed.), Butterworth-Heinemann, Oxford, UK, 1976.

[10] L. D. Landau, E. M. Lifshitz, The Classical Theory of Fields. Vol. 2 (4th ed.), Butterworth-Heinemann, Oxford, 1975.

[11] R. P. Feynman, A. R Hibbs, Quantum Mechanics and Path Integrals, (McGraw-Hill, NY, 1965.

[12] L.M. Martyushev, V.D. Seleznev, Maximum entropy production principle in physics, chemistry and biology, Physics Reports 2006, 426, 1 - 45.

[13] I. Gyarmati, Non-Equilibrium Thermodynamics: Field Theory and Variational Principle, (Springer, NY, 1970), pp. 10-15.

[14] M. Katz, J. Reimann, An Introduction to Ramsey Theory: Fast Functions, Infinity, and Metamathematics, Student Mathematical Library; American Mathematical Society: Providence, RI, USA, 2018; Volume 87, pp. 1–34.

[15] R. Graham, S. Butler, Rudiments of Ramsey Theory, 2nd ed.; American Mathematical Society: Providence, RI, USA, 2015; pp. 7–46.

[16] Y. Li, Q. Lin, Elementary Methods of the Graph Theory, Applied Mathematical Sciences; Springer: Cham, Switzerland, 2020; pp. 3–44.

[17] Gray C. G., Karl G., Novikov V. A. From Maupertuis to Schrödinger. Quantization of classical variational principles. Am. J. Phys. 1999, 67, 959–961.

[18] ] Gray C. G., Karl G., Novikov V. A., The Four Variational Principles of Mechanics, Annals of Physics, 1996, 251 (1), 1-25.

[19] Landau L. D., Lifshitz E.M., Quantum mechanics: non-relativistic theory, Volume 3 of Course of Theoretical Physics, 3$^{rd}$ Ed., Pergamon Press, Oxford, UK, 1991.

[20] Cremaschini C., Tessarotto M. Unconstrained Lagrangian Variational Principles for the Einstein Field Equations, Entropy 2023, 25(2), 337.

[21] Frenkel M., Shoval Sh., Bormashenko Ed. Fermat Principle, Ramsey Theory and Metamaterials, Materials 2023, 16(24), 7571.

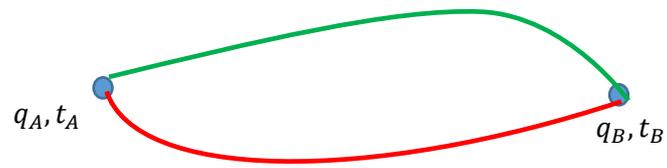

**Figure 1**. Actual (colored with red) and trial (colored with green) pathways of the system are depicted.

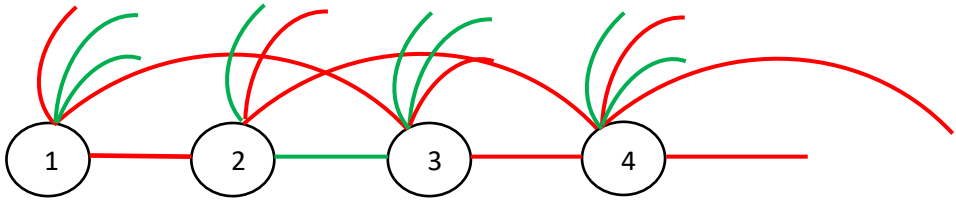

**Figure 2**. Infinite number of points $q_1, q_2 \ldots q_N \ldots$ in the configurational space of the physical system. The points represent the vertices of the infinite graph. The vertices are connected with a red link, when the variational principle governing the behavior of the physical system is fulfilled; according to the Ramsey wording these points are "friends"; red links correspond to actual trajectories of the physical system; the vertices are connected with a green link, depicting the trial trajectories, frustrating the variational principles; these points are strangers. Infinite monochromatic clique will necessary appear in the graph.